\documentclass[conference]{IEEEtran}
\IEEEoverridecommandlockouts
\usepackage{tikz}  
\usepackage{cite}
\usepackage{amsmath,amssymb,amsfonts}
\usepackage{algorithmic}
\usepackage{graphicx,caption}
\usepackage{textcomp}
\usepackage{xcolor}

\usepackage{url}  
\usepackage{color,soul}
\usepackage{enumitem}
\usepackage{supertabular}
\usepackage{booktabs}

\usepackage{tikz}
\usepackage{textcomp}
\usepackage{hyperref}
\usepackage{lipsum}

\newcommand\copyrighttext{%
	\footnotesize \textcopyright 2012 IEEE. Personal use of this material is permitted.
	Permission from IEEE must be obtained for all other uses, in any current or future
	media, including reprinting/republishing this material for advertising or promotional
	purposes, creating new collective works, for resale or redistribution to servers or
	lists, or reuse of any copyrighted component of this work in other works.
}
\newcommand\copyrightnotice{%
	\begin{tikzpicture}[remember picture,overlay]
		\node[anchor=south,yshift=10pt] at (current page.south) {\fbox{\parbox{\dimexpr\textwidth-\fboxsep-\fboxrule\relax}{\copyrighttext}}};
	\end{tikzpicture}%
}

\usepackage{subfiles} 

\def\BibTeX{{\rm B\kern-.05em{\sc i\kern-.025em b}\kern-.08em
		T\kern-.1667em\lower.7ex\hbox{E}\kern-.125emX}}

\setlist[itemize]{noitemsep, topsep=0pt}
\def\rot{\rotatebox{45}}

\begin{document}
\title{Measuring source code conciseness across programming languages using compression}

\author{\IEEEauthorblockN{Lodewijk Bergmans, Xander Schrijen, Edwin Ouwehand \& Magiel Bruntink}
		\IEEEauthorblockA{\textit{Software Improvement Group},
			Amsterdam, The Netherlands \\
			\texttt{\{l.bergmans, x.schrijen, e.ouwehand, m.bruntink\}@softwareimprovementgroup.com}
		}
}

\maketitle           

\copyrightnotice

\begin{abstract}
It is well-known, and often a topic of heated debates, that programs in some programming languages are more concise than in others.
This is a relevant factor when comparing or aggregating volume-impacted metrics on source code written in a combination of programming languages.
In this paper, we present a model for measuring the conciseness of programming languages in a consistent, objective and evidence-based way.
We present the approach, explain how it is founded on information theoretical principles, present detailed analysis steps and show the quantitative results of applying this model to a large benchmark of diverse commercial software applications.
We demonstrate that our metric for language conciseness is strongly correlated with both an alternative analytical approach, and with a large scale developer survey, and show how its results can be applied to improve software metrics for multi-language applications.
\end{abstract}


\section{Introduction}

Both in the academic literature and in daily practice there are a lot of claims and opinions about the differences between programming languages. The term `programming language wars' has even been coined to refer to the ``broad divergence and impact of language designs, including often pseudo-scientific claims made that they are good or bad" \cite{Stefik2014}.
One particular topic of discussion \---with more opinions than evidence\--- is the suitability, expressiveness, or conciseness of programs in different languages (e.g. \cite{Graham2002,Prechelt2000a,SoftwareProductivityResearch2007}).

\subsection{Motivation}
The amount of source code of systems, components or modules plays a large role in many metrics, human judging and decision making.
The main reason for this is that the amount of source code in a software entity serves as a first indicator of e.g. the amount of functionality, or the effort for creating or maintaining such an entity.

In software engineering, many software metrics are used that are influenced by the amount of source code, and hence by the conciseness of that source code.
For example, when identifying duplicated code, not only the amount of duplicates, but also the amount of code that is duplicated is an important factor. Or when identifying `god classes' \cite{riel1996object}, the size of the god class is a relevant indicator for the severity of the problem. Also, at an architecture level, understanding the volume of components can be crucial for understanding the impact of issues and decisions about those components.

However, comparing the source code from different programming languages based on the number of characters or the lines of code does not do justice to both the amount of source code or the amount of work involved in that source code.

This is relevant since most modern software systems are developed using multiple programming languages.
A typical example of this are different languages used for front-end development (e.g. JavaScript), business logic (e.g. Java), and data management (e.g. SQL).
As a result, the application of software metrics to systems or portfolios of systems requires that the differences in conciseness between languages is taken into account.

In this paper we will investigate measures for language conciseness, not with the intention of mingling into the language expressiveness discussions, but rather with the intention of enabling software metrics that are meaningful even when they cover source code in multiple programming languages.

\subsection{Problem statement}
The problem statement for the work we present in this paper is as follows:
\\
\emph{Given that:}
\begin{itemize}
    \item We are interested in measuring source code properties (calculating software metrics) at a system- and portfolio scale to provide insight into the quality characteristics of software.
    \item System- and portfolio-level metrics require the aggregation of metrics over multiple files, components and systems, written in a wide range of programming languages.
    \item Comparing the source code volume in different programming languages based on the number of characters or the lines of code does not do justice to both the amount of work involved or the amount of functionality in that source code.
\end{itemize}
\emph{How can we measure and compare the conciseness of source code in different languages?}


\subsection{Context and constraints}
The work presented in this paper has been motivated by a practical need: at the Software Improvement Group, we run software metrics on the source code for a wide range of software applications from multiple development organizations. With currently close to 50 million lines of code analysed daily, this has to be a fully automated process (after an intake period). Many of these applications involve multiple languages, and our analysis tools can currently handle over 250 different programming languages.

One of the key elements of this source code analysis is measuring the maintainability of software applications (according to the methodology documented in \cite{Baggen2012,Heitlager2007,Visser2016}), often for a portfolio of customer systems, and repeatedly over time.
To properly aggregate, interpret and compare the maintainability metrics, especially in relation to the size of systems and language parts of systems, the differences in conciseness of the involved languages must be taken into account.

Hence the following constraints and considerations apply to our work:
\begin{enumerate}
\item The approach must be \textbf{applicable to all (text-based\footnote{
		Indeed, the restriction to text-based languages is a limitation, but visual and low-code programming languages have too many practical hurdles in obtaining a representative source code representation that is free from generated template structures.})
	programming languages} we encounter. Developing and maintaining a specific semantic analysis and/or parser may not be feasible for each of these languages.
\item We aim for a \textbf{fact- or measurement-based approach}, that avoids the subjective opinions of developers (as these are known to vary wildly, and very sensitive to personal experiences and preferences). So we want to base the analysis on a source code benchmark or other data sources.
\item The approach must be \textbf{repeatable} over time, possibly with a varying set of applications, while yielding comparable results.
\item Preferably, we are \textbf{not depending on external parties} for data sources or benchmarks.
\end{enumerate}
We are approaching this problem and handling these constraints from an \textit{engineering perspective}: this means we are for example willing to trade in accuracy, when necessary to come to reasonable, repeatable and practical measures.


This paper is organized as follows: the next section presents some more background, to lay the groundwork for the rest of the paper, and discusses related work.
In section \ref{ref:approach}, we present the detailed steps of our approach, and in section \ref{ref:data_analysis}, we discuss the data analysis and the metric aggregation approach. In section \ref{ref:validation}, we discuss several approaches to check the validity of our results. Finally, we discuss the application of our approach to software quality metrics, and threats to validity in section \ref{ref:discussion}, and summarize the contributions and some of the next steps in section \ref{ref:conclusion}.

\section{Background and Related Work}\label{ref:background}

\subsection{Setting the stage} \label{ref:setting_the_stage}
To disambiguate some commonly used terms, we first provide definitions for a few core concepts:
A \textit{program} is a collection of instructions that can be executed by a computer to perform a specific task \cite{enwiki:computer_program}.
A program is specified by \textit{source code}, following the rules of a \textit{programming language}. In this paper we consider text-based source code only; programs may be defined by one or more files with source code, and a single program may consist of source code in different programming languages.

A key concept behind our approach is to consider \textit{source code as an encoding of knowledge about the program, in a format that a compiler or interpreter can process}.
This knowledge spans multiple domains; from the knowledge about the programming language and its semantics, libraries that are used, coding guidelines, design decisions and design patterns, to knowledge about the requirements for the program and the application domain.
It is a developer's task to gather and structure that knowledge, and encode it into source code.

Accordingly, we propose the following definitions:
\begin{description}[topsep=0pt]
     \item[program conciseness] is the amount of code that is required to express a given amount of information about a program. In other words: \textit{the ratio between the amount of information and the amount of source code for a program}.
     \item[language conciseness] is the typical ratio between the amount of information and amount of code in programs written in that language. The concrete definition of `typical ratio' will be a distinct topic in the remainder of this paper.
\end{description}
It should be clear that program conciseness is \textit{not only} influenced by the programming language: other influencing factors are the application domain, the experience and background of the developers, the programming style, and coding standards, among others.
The assumption is that when combining the program conciseness for a sufficient number of programs in a given language, this can be used to calculate an adequate typical conciseness indicator for that language.

Next, we will present some background on measuring information content, and then discuss related work on language conciseness.

\subsection{On measuring information content}
Above we used the term  `amount of information' to define conciseness: this is not a common metric, and may seem extremely hard to measure. However, there exists both solid theory about measuring information in a dataset (and  source code is an example of such a dataset), and tested practical approaches that apply that theory.

The \textit{Kolmogorov complexity}\cite{Kolmogorov1963,Li2008} of a dataset can be described as the smallest size that a dataset can be compressed to\footnote{
   The more accurate definition is: the Kolmogorov complexity of an object is defined as the length of the shortest computer program (in a predetermined programming language) that produces the object as output. We will not need to use this information-theoretic definition in this paper, since we focus on the practical approximation through compression algorithms.
}.
Kolmogorov complexity is theoretically incomputable, but can be approached from above with compression algorithms \cite{Li2008}. For example, it has been proven that Ziv-Lempel complexity\cite{Ziv1977} or similar techniques can, assuming infinite computing power and memory, compress any object as close as possible to the theoretical minimum compressed size, without any additional knowledge about the object\cite{Juola2008b}.
Indeed, there are many successful practical applications of Kolmogorov complexity that use compression algorithms.
Examples are: the comparison of (evolution of) objects\cite{Li2004}, clustering\cite{Cilibrasi2004} and causal inference\cite{Budhathoki2017}. Hence, it is safe to claim that the information content of a dataset, such as source code (which lends itself quite well to compression) can be approximated to a level that may have practical applicability.

\subsection{Related work}\label{ref:relatedwork}
Our prime motivation for measuring language conciseness is to do justice to the differences in conciseness between --source code in different-- programming languages. This has been considered in its most principled form in the context of counting lines of code. In that context, the notion of \textit{Logical Source Lines Of Code} has been coined, which aims to count statements, rather than \textit{Physical Source Lines Of Code}; already a first abstraction from the physical representation of source code towards a more semantic measurement of volume.
However, Logical Source Lines of Code does not do justice to the differences among programming languages, and how expressive statements in different languages are (or can be).
Also, note that counting Logical Source Lines of Code requires parsing of the language, and hence a grammar for each language.
Although the required result of this parsing is only a simple identification of statements, language features like comments and string literals must be properly handled, or the results can be completely off.
Building (full) parsers can be hard and expensive, e.g. parsing C and C++ is notoriously difficult  \cite{Padioleau2009}, as a result of macro's and preprocessing.

An important category of related (but not equivalent) work, are the language-specific function points to lines of code factors; also called \textit{language gearing factors}.
Functions points\cite{albrecht1983fp} is a technique that aims to estimate the size of software as a reflection of the amount of functionality a software application implements.
The function points method does not require source code, but only a requirements specification, so it can already be applied at a stage where there is no software yet (although that applies more to waterfall development approaches).
A related approach called Automated Function Points \cite{omg2014afp}, is based on source code analysis, but still requires substantial manual input before conducting the automated process, which makes it unsuitable for large scale benchmarking.

A language gearing factor thus expresses how many statements in a given programming language are needed to express a certain amount of functionality (expressed as a function point).
Hence, a language gearing factor is a measure of the expressiveness, or conciseness of the language.
Language gearing factors, however, are very sensitive to reuse, both within the application, and reuse from external libraries.

Examples of sources for language gearing factors are \cite{QSM2014,spr2007,Jones2017}; the reported factors are based on comparing function point estimates for a benchmark of systems with the size of their source code, or by qualitative judgement of language properties\cite{spr2007}.
Of these, the SPR table has been the most elaborate source, covering a wide range of languages. However, some of its language gearing factors are rather un-intuitive (e.g. when comparing two very similar languages with rather different factors), and it has not been updated since 2007, hence excluding many modern (and popular) languages\footnote{
	At the Software Improvement Group, this has been the main driver for the reported research}.

The notion of 'language conciseness' has a clear intuitive relation to the 'expressiveness' of programming languages.
In this area, \cite{Felleisen1991} is an important paper that discusses a formal framework for reasoning about the expressiveness of programming languages.
This is entirely based on the formalization of language semantics, which is not applicable for the practical application we aim for in this paper.
The paper does contain the following relevant statement: ``By studying a number of examples we have come to the conclusion that programs in less expressive languages exhibit repeated occurrences of programming patterns and that this pattern-oriented style is detrimental to the programming process"; this does support our assumption that less expressive languages will be both less concise and require more time to program the same amount of functionality.

A final category of related work aims at identifying the conciseness or expressiveness of programming languages through empirical investigations: Berkholz\cite{Berkholz} investigates whether the typical size of source code commits could be an indicator of programming language expressiveness, and provides a ranking of language expressiveness based on the median commit size per language for circa 50 languages. The \textit{Hammer Principle}\cite{MacIver2017} is a large scale survey among programmers to identify a ranking of programming languages with respect to properties such as, among many others, expressiveness, verbosity and conciseness. The \textit{Hammer Principle} survey results are also used in \cite{Meyerovich2012} to create rankings of various language properties. The Berkholz and \textit{Hammer Principle} studies provide a highly relevant tool for us to validate whether our results conform to both the practices and the intuition of developers; if this is the case, it makes our results more useful (because more relatable).

In \cite{Prechelt2000a}, an empirical investigation is conducted, where a single, small, program, was implemented in seven different languages, by 80 (student) developers. This results in comparison of the programming languages with respect to e.g. program size and programming effort.
Finally, both \cite{wolfram2014} and \cite{Nanz2015} conduct a comparison of programming languages based on the \textit{Rosetta Code} repository, which contains implementations of a set of programming tasks in a wide range of programming languages; this allows for comparing programs for the same task\footnote{Although often written with different interpretations or different goals, such as brevity or readability.}, written in multiple programming languages.

Related work on approaches to measuring language conciseness that are similar to this paper:
In the domain of natural languages, research has been conducted on determining the `linguistic complexity' of natural languages, based on the `relative informativeness' of text samples by adopting file compression to estimate Kolmogorov complexity \cite{Juola2008b,Szmrecsanyi2016}.

The first version of our work was published within a Ph.D. thesis \cite{Raemaekers2015}, where \textit{relative }language conciseness was calculated based on compression ratios of source code in those languages. There it was applied to calculate the size of applications when rebuilding them in another language.
Subsequently, in the M.Sc. thesis of Ouwehand\cite{Ouwehand2018msc}, we investigated what the most suitable algorithms are for compressing source code of software systems, and how to build a benchmark for comparing their compression ratios.

This paper is a significant improvement over these previous works by adopting a much larger benchmark, covering more languages, presenting a novel aggregation method to calculate language conciseness, and a quantitative validation of the results with two other sources.

\section{Approach}\label{ref:approach}

In this section, we will describe our benchmark-based approach to determine language conciseness, which is objective, repeatable, and applicable to any (text-based) programming language.

In section \ref{ref:background}, we already outlined the basic premises underlying our approach; the key element is to view all source code of a program, grouped per language, as data for which we can estimate the information content through data compression. By comparing the information content to the original size, a \textit{compression ratio} is determined that is an indication of the conciseness (or verbosity) of the program. We will apply this analysis to a large benchmark of applications to aggregate to a representative \textit{language conciseness} measure.

As the source for the benchmark, we have used an in-house software warehouse, which contains the source code of over 5000 curated industrial software applications, covering more than 280 languages (as of mid 2021), that have been scoped and analysed individually over the years.

For collecting our language conciseness benchmark, we take the following steps:
\begin{enumerate}[topsep=0pt]
\item \textbf{Gather snapshots}: our software warehouse consists---for a large share of the systems---of many snapshots per system, over time. For each system, we use the latest version that is in the software warehouse.
\item \textbf{Select and split each system in its language parts}: to ensure that we focus on actual production source code, the files for each snapshot are scoped, which among others excludes generated code and test code. After this, the files in the system are categorized according to their language technology (or excluded if they are pure data, such as image files).
\item \textbf{Stripping comments and white-space}: to avoid the influence on compression results of natural language in source code comments, we strip all comments---and white-space---from the source code, using the \texttt{cloc} tool\footnote{ \texttt{github.com/AlDanial/cloc} }.
\item \textbf{Archiving and compression}: for each language part of a system snapshot, all source code is combined in a single tar file, which is then compressed by the LZMA2 compression algorithm, using the \texttt{xz} tooling\footnote{
   As obtained from \texttt{tukaani.org/xz/}.
   For the results in this paper, we have used version 5.1.0alpha, with compression parameters \texttt{--lzma2=preset=9e,lc=4,pb=0}.
}. An elaborate analysis of the various compression algorithms, and their ability to provide strong compression and consistency when compressing source code, has been conducted in \cite{Ouwehand2018msc}; LZMA2 was selected as the most suitable algorithm for this purpose.
We use a tar archive because it allows us to compress the archived files a whole; in a zip archive, where each file is compressed separately, we would not detect redundancy across files. The latter is especially interesting since it is related to the ability of a language to achieve modularity and reuse across files.
Section \ref{ref:threats} discusses some additional details about the choice of compression algorithms and usage of tar files.
\item \textbf{Aggregating compression results to language conciseness measures}: the previous step provides both the original size of the source code, and the compressed size: for both cases we use the number of bytes in the files; the result is a compression ratio (CR) between the original and compressed size.
Next, the compression ratios for the same language from all systems are aggregated into a single language compression ratio.
\end{enumerate}

We will discuss the aggregation algorithm in the next section, after showing some initial results.

\section{Data Analysis}\label{ref:data_analysis}

In this section we discuss the results of the data analysis we have performed; the characteristics of the data, and some additional design decisions we made based on the findings.

Before the analysis we excluded all visual-, process- and low code-languages---i.e. languages with no text-based source code.
During the analysis some systems failed to process correctly, for various implementation-related reasons.
Finally, we ended up with a subset of the benchmark that had successfully passed all processing steps, consisting of 4954 systems.
The size of the systems follows a log-normal distribution, with an average system size of 230 KLOC, and a median system size of 30 KLOC.
The dataset contains 103 different languages\footnote{
	Apart from the removal of visual-, process- and low code-languages, also a substantial amount of infrequently used languages did not make it to the final set, mostly because we did not include the oldest parts of our archives, or they were part of snapshots that were excluded when inconsistencies in the archive (meta-)data were found.}, with an average of 3.3 languages used per system.
The median number of data-points (we divide each system into distinct parts for each language; each of these parts becomes a data-point) \textit{per language} is 24, with many data-points (over 400) for the top-ten most popular languages.

After splitting the source code for each system into its language parts (i.e. the part that collects all source code in one specific language), we obtain a benchmark with 16,314 different data-points for our analysis.
Next, we take several steps of data cleaning:
\begin{enumerate}[topsep=0pt]
    \item \textbf{Remove extreme outliers}: here we remove extremely large language parts (more than 100 million characters), extremely small language parts (less than 10 lines of code), compression ratios that are less than 1, and that are over 100 (NB: note that a compression ratio \textit{x} means there is a ratio of 1:\textit{x} of the compressed to the original size.).
    \item \textbf{Remove CR outliers}: We remove all Compression Ratio (CR) values that are more than 2 Standard Deviations from the mean. A typical cause of outlier values are (very) large duplications in the source code (e.g. identical files, folders, components or libraries), which yield high compression ratios, but are a result of design decisions (or sometimes lack thereof) by the developers rather than language characteristics.
    \item \textbf{Remove smallest language parts}: We have made the design decision to remove 10\% of all the smallest language parts for each language. The motivation for this is two-fold: (1) Compression algorithms naturally work less well for small datasets, if only because there is less redundancy and repeating patterns: in other words, we obtain less information about conciseness from small program parts. (2) This is especially relevant since we are interested mostly in using the results for medium-sized programs (not toy examples or small projects), and a strong influence of small language parts (which typically have a much lower compression ratio because there is inherently less redundancy in small programs). Note that a fixed size threshold for all languages would not work out well, since for some categories of languages, e.g. for shell scripts, the typical size is much smaller than for other languages, e.g. those that are often used to build enterprise systems.
\end{enumerate}

\noindent
Table \ref{tbl:cleaning} shows the impact on these cleaning steps on the size of the dataset, and on the number of distinct languages remaining in the dataset.
\begin{table}
  \caption{Results after each step of the data cleaning process}\label{tbl:cleaning}
  \centering
\begin{tabular}{ l l r r }
\toprule
\textbf{Step} & \textbf{Description} & \textbf{\#data-points} & \textbf{\#languages}\\
\midrule
0.  &  After processing  &  16,314  &  103 \\
1.  &  Remove extreme outliers  &  15,687  &  102 \\
2.  &  Remove CR outliers  &  15,017  &  102 \\
  3.  &  Remove smallest language parts  &  13,476  &  94 \\
\bottomrule
\end{tabular}
\end{table}
\begin{figure*}
	\captionsetup{width=.9\textwidth}
	\begin{center}
		\includegraphics[width=0.9\textwidth]{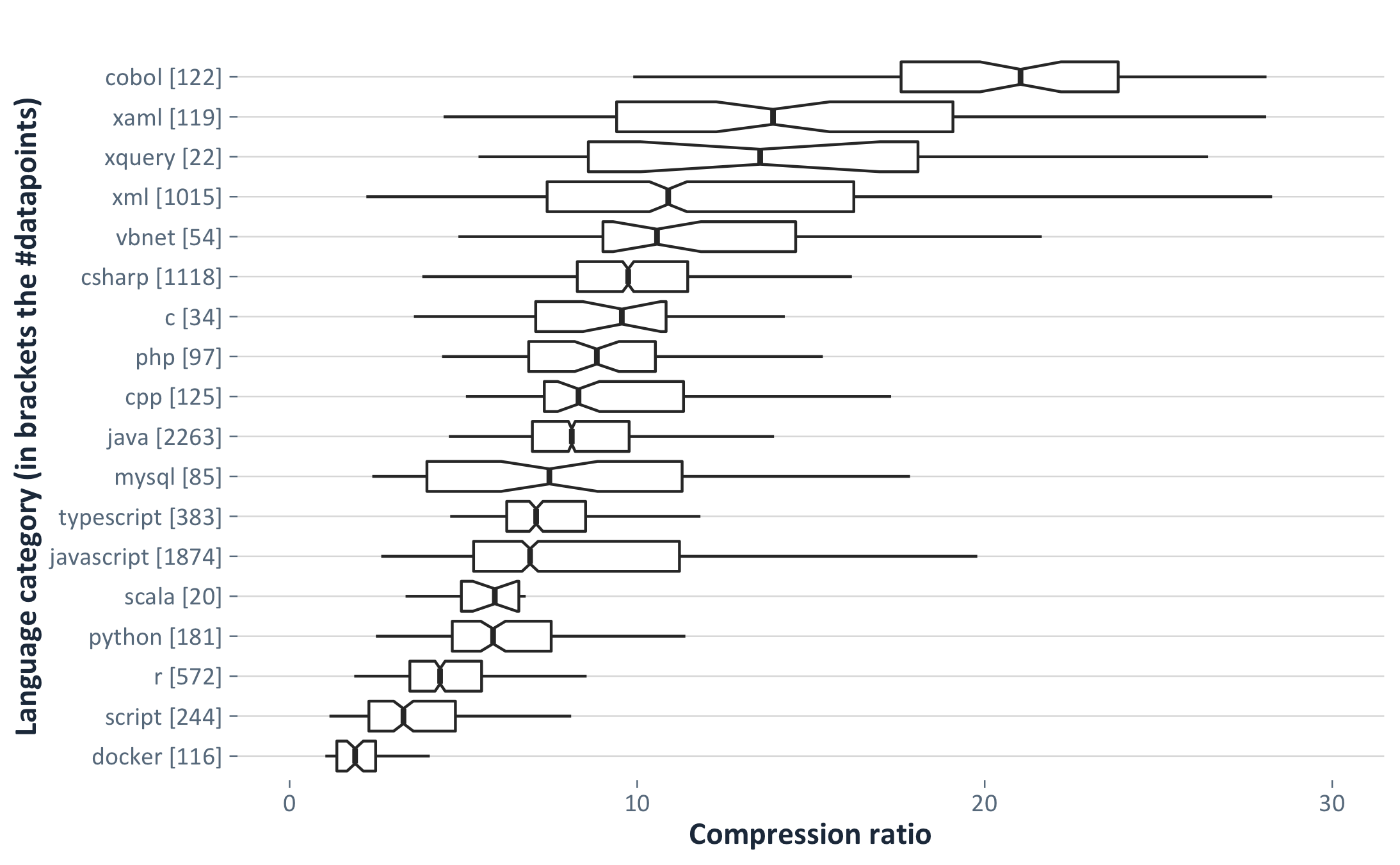}
		\caption{Box-plot of compression ratios for 18 frequently occurring languages in our benchmark. The languages are ordered by the \textit{median CR values}. The notches define a 95\% confidence interval for comparing the medians. } \label{fig:cr_boxplot}
	\end{center}
	\vspace{-15pt}
\end{figure*}
Figure \ref{fig:cr_boxplot} shows box-plots of the data for a selection of languages (mostly based on how common or well-known they are): the compression ratios shown are the ratio between the compressed and the uncompressed size of the individual language parts.
For each language, the number of data-points is shown in brackets.
This figure shows that the CR values are distinctly different between languages: median CR values vary from circa 4 for shell scripting languages, to around 8--10 for languages such as Java and C\#, to 20-25 for XML-based languages\footnote{
	Although especially XML itself is strictly not a programming language, it is included because developers often use XML documents for configurations or specifications that are an essential part of the program behaviour: XAML, XSLT and XSD source code are good examples of such XML-based specification languages.}.
The box-plot notches, which indicate the 95\% confidence interval for comparing the medians, show that for many languages, the median CR is distinctively different. It is expected that some languages are almost equally concise, and hence have similar medians.
On the other side, for some languages the range of CRs is very similar, and although it differs quite a bit between languages, there is also a wide variation in CR values between programs, especially for the languages with a high median CR.

To provide some more insights into the compression ratios, and illustrate some further design decisions, we show the CR data for a single language, for which we have selected \textit{Typescript}; this is a medium `popular' language with 383 data-points in our set, after all cleaning steps.
\begin{figure*}
	\captionsetup{width=.9\textwidth}
	\begin{center}
		\includegraphics[width=0.9\textwidth]{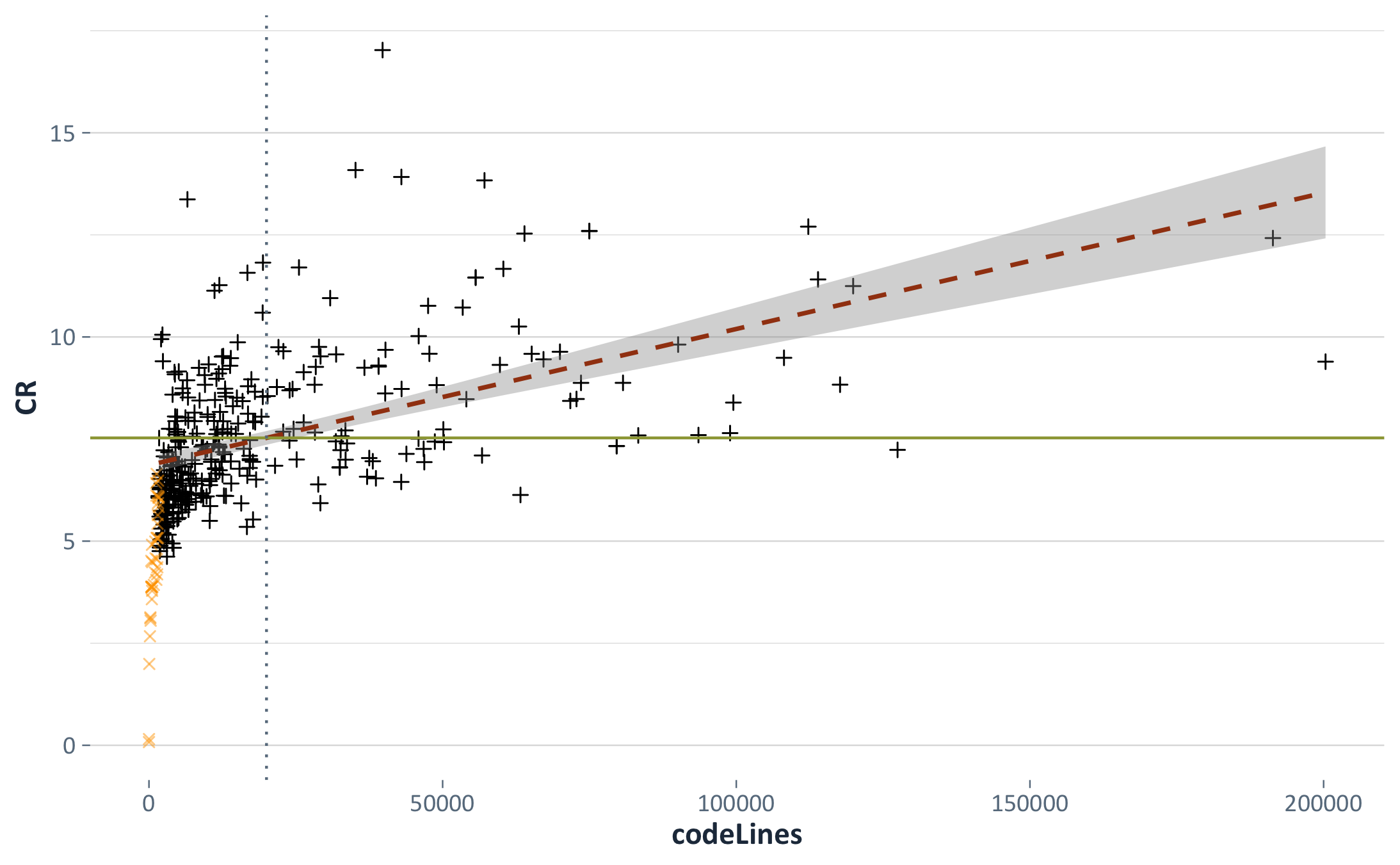}
		\caption{A plot of all data-points for Typescript. The orange crosses are the data-points that have been removed during cleaning. The red dashed line shows a linear regression with 95\% confidence interval. The grey dotted vertical line indicates the mean code-size, and the solid green line marks the calculated CR (7.5) for this language. } \label{fig:codesizeVsCR}
	\end{center}
	\vspace{-15pt}
\end{figure*}

Figure \ref{fig:codesizeVsCR} shows the data-points for the \textit{Typescript} language:
for this language there were almost no outliers (marked as orange crosses), only the 10\% smallest parts have been removed. The linear regression line shows that on average, larger programs are easier to compress.

One of the challenges is to determine a single CR that can be considered representative for Typescript source code.
Again, one of the criteria is that the selected value is particularly representative for the mid-range sized programs.
We have selected the following approach to aggregate to a single CR value:
\begin{enumerate}[noitemsep,topsep=0pt]
   \item First we define a linear regression line $Lr$ over all data-points for that language, where the x-values represent code size (LOC) and the y-values the CR: this is the red dashed line in figure \ref{fig:codesizeVsCR}.
   \item Then we select the y-value on that line at the mean code size (this corresponds to the dotted grey vertical). That y-value represents the CR for this particular language (it is marked by the solid green line).
\end{enumerate}
\noindent
We needed to define a threshold for the amount of data-points that is acceptable before adopting a compression ratio value.
This threshold is a balance between consistency and the number of technologies for which can we produce results.
After conducted a sensitivity analysis to test how many data-points are needed before obtaining a relatively stable CR value, we have observed that---at least for most languages---already after 10 data-points, the aggregated CR is very similar to the final version as derived from the full set of data-points for that language.

During the analysis of the data, we noticed that there are clear trends in the CR over time; for a majority of the languages, there was a repeated yearly incline or decline of the CR of 2-5\%\footnote{
	We do not have exact explanations of the reasons for these trends, but the evolution of languages and libraries can at least partially explain such effects.}.
Given that observation, we prefer to focus on more recent data; for the final version of the compression ratios, we reduce the set of data-points by removing, per language, the data-points that are older than 5 years, \textit{as long as} there are at least 100 data-points for that language.

Table \ref{tbl:CRoverview} shows the resulting representative compression ratios for the same set of languages as in Figure \ref{fig:cr_boxplot}.
After all filtering steps and application of thresholds, the total list with language compression ratios has 58 entries. CR values range from 2.2 for Docker, to 20.9 for Cobol.

\setlength{\tabcolsep}{2pt}
\begin{table}[h]
\caption{The final Compression Ratios for a selection of languages}\label{tbl:CRoverview}
	\centering
	\begin{tabular}{rlrrr}
		\toprule
		\#  & language\quad\quad\quad & n & median & CR \\
		\midrule
1 & cobol & 72 & 21.0 & 20.9 \\
2 & xaml & 88 & 13.9 & 14.6 \\
3 & xquery & 19 & 13.5 & 13.5 \\
4 & xml & 670 & 10.9 & 12.4 \\
5 & vbnet & 48 & 10.6 & 11.9 \\
6 & csharp & 994 & 9.7 & 10.2 \\
7 & cpp & 111 & 8.3 & 9.8 \\
8 & c & 28 & 9.6 & 9.3 \\
9 & php & 82 & 8.8 & 9.3 \\
	\bottomrule
\end{tabular}
\quad
\begin{tabular}{r l r r r}
	\toprule
	\#  & language & n & median & CR \\
	\midrule
10 & java & 2028 & 8.1 & 8.8 \\
11 & mysql & 70 & 7.5 & 8.7 \\
12 & javascript & 1647 & 6.9 & 8.5 \\
13 & typescript & 342 & 7.1 & 7.5 \\
14 & python & 161 & 5.9 & 6.5 \\
15 & scala & 17 & 5.9 & 5.6 \\
16 & r & 502 & 4.3 & 4.8 \\
17 & script & 194 & 3.3 & 4.3 \\
18 & docker & 57 & 1.9 & 2.2 \\
		\bottomrule
	\end{tabular}
\end{table}

\section{Validation}\label{ref:validation}

\begin{table*}[ht]
	\vspace{-8pt}
	\caption[caption]{Spearman rank correlations between our CR values, the Berkholz ranking, \\and 6 different characteristics from the Hammer Principle developer survey\footnotemark .}\label{tbl:correlations}
	\centering
	\begin{tabular}{rrrrrrrrr}
		& \rot{\textbf{CR}} & \rot{\textbf{Berkholz}} & \rot{\textbf{Verbose}} & \rot{\textbf{Terse}} & \rot{\textbf{Low-level}} & \rot{\textbf{high-level}} & \rot{\textbf{Expressive}} & \rot{\textbf{Acc. compl.}} \\
		\toprule
		\textbf{CR} & 1.00 & 0.68 & -0.84 & 0.82 & -0.63 & 0.55 & 0.74 & -0.59 \\
		\midrule
		\textbf{Berkholz} & 0.68 & 1.00 & -0.62 & 0.67 & -0.58 & 0.61 & 0.85 & -0.49 \\
		\midrule
		\textbf{Verbose} & -0.84 & -0.62 & 1.00 & -0.95 & 0.69 & -0.63 & -0.77 & 0.64 \\
		\textbf{Terse} & 0.82 & 0.67 & -0.95 & 1.00 & -0.65 & 0.57 & 0.86 & -0.62 \\
		\textbf{Low-level} & -0.63 & -0.58 & 0.69 & -0.65 & 1.00 & -0.89 & -0.66 & 0.72 \\
		\textbf{High-level} & 0.55 & 0.61 & -0.63 & 0.57 & -0.89 & 1.00 & 0.73 & -0.70 \\
		\textbf{Expressive} & 0.74 & 0.85 & -0.77 & 0.86 & -0.66 & 0.73 & 1.00 & -0.65 \\
		\textbf{Acc. compl.} & -0.59 & -0.49 & 0.64 & -0.62 & 0.72 & -0.70 & -0.65 & 1.00 \\
		\bottomrule
	\end{tabular}
\end{table*}

\footnotetext{The number of matching languages for the correlations is 20 between compression ratio's and the Hammer Principle table, and 15 between the Berkholz ranking and the other data. The p-values for the above correlation matrix are all below 0.062; the correlations between CR and the Hammer data all have a p-value below 0.012, the correlation between CR and Berkholz has a p-value of 0.003, and the correlations between Berkholz and the Hammer data have p-values below 0.062. In other words, these correlations are quite strong and significant, even though there is only a limited set of languages overlapping between the datasets.}

When manually inspecting table \ref{tbl:CRoverview}, at least two observations can be made:
\begin{enumerate}[noitemsep,topsep=0pt]
	\item The ordering of the languages (from high CRs, i.e. verbose languages, to low CRs, i.e. concise languages), does roughly follow the intuition that most developers will have: Cobol and XML variants at the top, then general purpose languages such as VB, C\# and Java, followed by scripting languages such as JavaScript and Python, and functional languages such as Scala and R, followed by shell scripts at the most concise end of the spectrum.
	\item Our approach results in (aggregated) CR values that are quite distinct from the median CR for that language in many cases. That might seem to imply that the aggregated CR is not very representative for the underlying data, but tests have shown that in fact it \textit{is} quite representative: we compared our aggregation approach to several alternatives (including mean and median values); when comparing, for each data-point from our benchmark, the \textit{Root-Mean-Squared deviation} between the calculated CR and the CR at individual data-points, our CR aggregation gave the lowest deviation, when averaging over all languages. Also the deviations had the least amount of variation across languages for our selected aggregation method. Due to lack of space, we cannot show the details of the analysis in this paper. 
\end{enumerate}

\noindent
In section \ref{ref:relatedwork}, we discussed the Berkholz\cite{Berkholz} and \textit{Hammer Principle}\cite{MacIver2017} studies, which both provide relevant language rankings.
The Berkholz ranking attempts to provide a ranking of language expressiveness; although expressiveness is not equivalent to conciseness, it is a safe assumptions that expressiveness and conciseness should correlate: in other words, that a more expressive language is more concise (and will thus have a lower CR value).
From the Berkholz dataset, we use the ranking number of the languages to identify a possible correlation with the compression ratio for those languages.
Unfortunately, only 15 of the languages in the Berkholz study appear also in our final list of languages.
We discuss the results below.  

The \textit{Hammer Principle} study is a survey with many\footnote{
	The web page mentions ``Based on responses from 398469 people, ...'', although \cite{Meyerovich2012} mentions 13,000 in the first 2 years.
} respondents, requesting the opinion of the respondents on many aspects of a programming language (100 questions).
The survey covers 49 languages.
We have selected the following six questions as relevant to the conciseness of a language:
\begin{enumerate}[noitemsep,topsep=0pt]
	\item ``Code written in this language tends to be \textbf{verbose}.''
	\item ``Code written in this language tends to be \textbf{terse}.''
	\item ``This is a \textbf{low-level} language.''
	\item ``This is a \textbf{high-level} language.''
	\item ``This language is \textbf{expressive}.''
	\item ``There is a lot of \textbf{accidental complexity }when writing code in this language.''
\end{enumerate}
Considering these six different, but related, questions also provides the opportunity to check the internal consistency in the responses.
For each of these questions, all languages have been ranked from most to least 'matching' the questions (e.g. from most to least verbose). We are using the ranking number of each language for a correlation analysis with our CR values and the Berkholz ranking.

Table \ref{tbl:correlations} shows the results of the Pearson rank correlation; generally, this analysis shows strong positive and negative correlations, that are also internally consistent (e.g. verbose and terse languages have a very strong negative correlation, and similar for high-level and low-level languages).
Hence we can conclude that our CRs correlate strongly with the Berkholz language expressiveness ranking.
Our CRs also have a very strong positive correlation with language verbosity, as resulting from the Hammer Principle developer survey.

From the various analyses in this section, we can conclude that:
\begin{enumerate}[noitemsep,topsep=0pt]
	\item The aggregated CR values are representative for the data-points in our benchmark, and show a significant distinction in conciseness for the majority of the languages.
	\item The Compression Ratio of languages have a strong positive correlation (0.68) with the Berkholz programming language ranking, for the 15 languages that are overlapping both approaches. 
	\item According to the results from the Hammer Principle survey, the CR-based ranking of programming languages matches the perceptions of the developers quite well, especially \textit{verbosity }and \textit{terseness} ($  $conciseness) are very strongly correlated (-0.84 respectively 0.82); that also matches the intention that compression ratios are a measure of conciseness.  
	From this we conclude that, to a large extent, using CRs as an indicator of language conciseness is confirmed by (or consistent with) developer opinions.
\end{enumerate}

\section{Discussion and Threats to Validity}\label{ref:discussion}

\begin{figure}[b]
	\begin{center}
		\includegraphics[width=1.0\columnwidth]{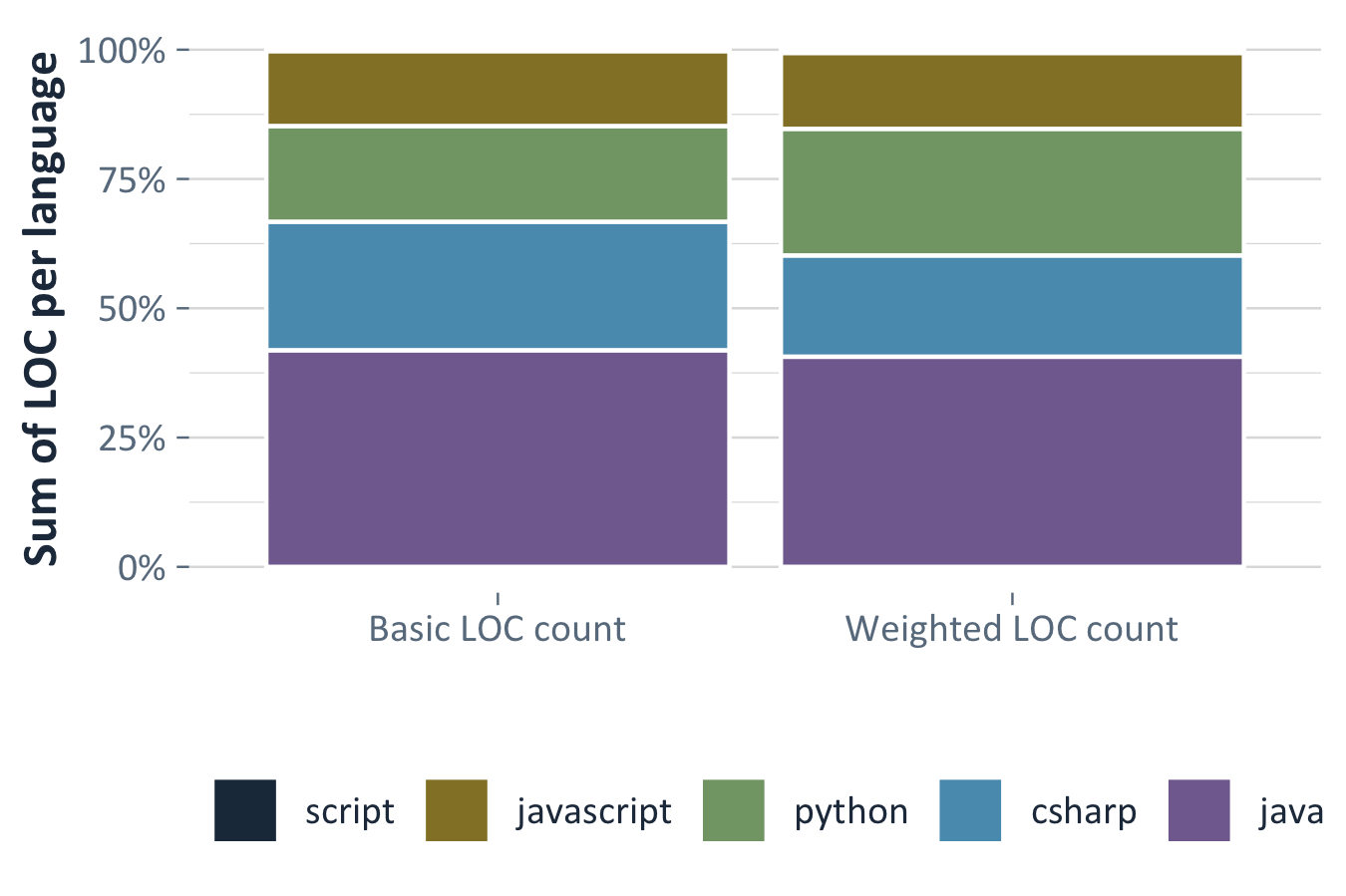}
		\caption{Proportions of language volumes in the system, based on basic, respectively weighted LOC counts. } \label{fig:ExampleLangDistr}
	\end{center}
	\vspace{-15pt}
\end{figure}


\subsection{Application to maintainability metrics}

To illustrate how our results can be applied, we show how compression ratio's can be used for weighting of code volumes, and how this can affect the results of metrics.

To show concrete results, we calculated source code metrics for a sample system, for which we have used the open-source ANTLR implementation\footnote{
	The source code can be found at \texttt{github.com/antlr/antlr4}, the version we used dates from April 19, 2019.
}.
The main reason for choosing ANTLR is that its implementation contains substantial parts in different languages, including Java, C\#, Python, JavaScript, and small amounts of Shell scripts.

Figure \ref{fig:ExampleLangDistr} shows in its left column the proportion of each of these languages in the system implementation, based on their respective Lines of Code.
The major segment is Java, followed by C\#, and two similar segments of Python and JavaScript.
The segment for Scripts is so small that it does not show in the bar chart.

However, as we have argued in this paper, a line of code in one language is not comparable to a line of code in another language.
In the right column of Figure \ref{fig:ExampleLangDistr}, we show the division after weighting the LOC count with the CR values of each language: weighting is done by dividing the LOC count by the compression ratio: so a line of code in a verbose language, with a high CR value, will weigh less, compared to a line of code in a more concise language, with a low CR.
For the ANTLR code, this shows especially that ---when applying weights--- the volume of Python code is at least as large as the volume of the C\# code, whereas the reverse conclusion would be drawn from the unweighted line counts.
The intuitive meaning of volume based on lines of code that are weighted by CR is that it corresponds to the amount of information in that source code.
\begin{figure}
	\begin{center}
		\includegraphics[width=1.0\columnwidth]{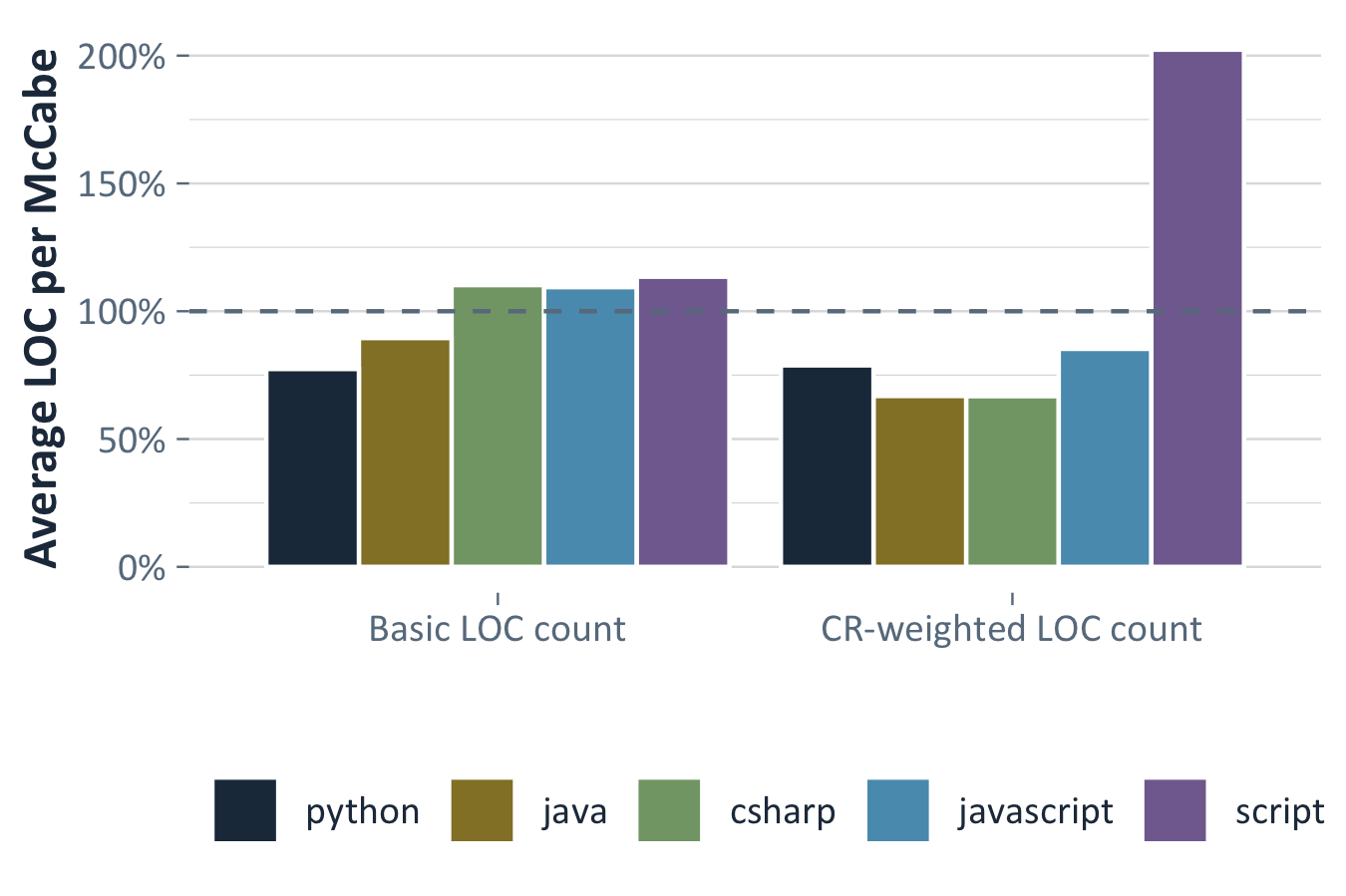}
		\caption{Relative LOC per McCabe values for different languages, comparing the basic LOC counts on the left, with the CR-weighted LOC counts on the right. } \label{fig:ExampleMcCabeWeighting}
	\end{center}
		\vspace{-15pt}
\end{figure}

The second example, based on the same ANTLR source code, is shown in Figure \ref{fig:ExampleMcCabeWeighting}: this shows the impact of CR-based weighting on source code metrics.
As an example, we have collected the ratio between McCabe value and lines of code in functions and methods.
Specifically, we collect the average of the number of lines of code per McCabe in each function, in other words, the average number of lines per decision point in the control flow.
To ease comparison, Figure \ref{fig:ExampleMcCabeWeighting} shows these values normalized, such that 100\% is the average LOC to McCabe ratio over all languages\footnote{
	This normalization makes it easier to compare the unweighted and weighted values.
}.

This example shows that source code metrics that are relative to the volume of the source code (which is the case for the majority of the metrics) may lead to different conclusions when applying CR-weighting of the LOC counts.
For example, in the unweighted version on the left side of Figure \ref{fig:ExampleMcCabeWeighting}, the average LOC : McCabe ratio is very similar for C\#, JavaScript and Shell scripts. But in the CR-weighted version, C\# has a lower LOC : McCabe ratio than JavaScript, and Shell scripts have a ratio that is almost double, caused by the conciseness of shell scripts (and correspondingly low CR): Shell scripts have a relative low amount of decision points per line of code, when taking their conciseness into account.
Also, in the unweighted version, the ratio is lower for Python than for Java, but in the weighted version this is reverse.

The examples in this subsection serve to illustrate the impact of CR-based weighting of LOC counts, and how this can lead to different insights.
We have used this straightforward approach for the purpose of illustration: for the aggregation of software metrics to system level, averaging of metrics is usually not suitable, because many software metrics follow a power law distribution.
We have adopted a structured approach to deal with the aggregation of software metrics such as proposed in \cite{Baggen2012} and \cite{Heitlager2007}, where metric values are first translated into risk categories, and then assigned a rating based on a benchmark.

Also, for presenting the weighted lines of code to developers, we do not directly use the lines of code divided by compression ratios in practice: that effectively compares source code based on the typical amount of information that it represents.
However, this is not a very familiar way of thinking about the size of a program for developers.
Hence, we map the volume of a program (as measured in for example lines of code) to a measure that corresponds to the amount of effort that is estimated to be required to develop that program: expressing volume as 'deduced build effort' is much easier to interpret.
The actual calculation we use and its underlying design principles are beyond the scope of this paper.


\subsection{Threats to Validity}\label{ref:threats}
We briefly discuss a number of (potential) threats to validity:
\begin{itemize}
	\item \textit{Compression algorithms are not perfect, hence they yield no perfect measure of information content}: the impact of this issue is limited because we do not use the absolute compression values per se, but use them to find a characteristic value that allows for \textit{relative comparison} of different languages. So imperfect compression is mostly a risk if compression is systematically different between programming languages.
	\item \textit{Different compression algorithms might yield different results}: Also here it is important to think of compression ratios not as absolute measurements, but as a relative comparison between programming languages. The compression ratios from different algorithms might be different, but they closely follow the same distribution. However, some compressors, such as \texttt{gzip} and \texttt{bzip2} compress data in blocks. These are unfit for purpose because no repetition is detected across block boundaries, which skews the results for larger systems. In \cite{Ouwehand2018msc} it has been shown that results vary by at most 1\% when using compressors that do not have this mechanism.
	\item \textit{Archive overhead}: A tar archive pads files with zeros to meet a certain block size. We risk skewing the result when compressing systems which are split into many smaller files. To combat this we calculate the compression ratio based on the size of the system before it is archived (so not the size of the tar archive). Because the padding compresses very well, it has no significant influence on the compressed archive size.
	\item \textit{Benchmark may not be representative} Using one specific benchmark always has its limitations: in this case, the benchmark consists mostly of industrial software systems, of varying size, but including many (very) large systems. The benchmark is somewhat skewed towards bespoke administrative- and information-systems developed in North-Western Europe. We would like to point out the importance of a benchmark with real-world projects of sufficient scale; there are several repositories with algorithm implementations in multiple languages, but these are mostly unsuitable for our purposes (mostly due to the impact of small programs on compressibility).
	\item \textit{Programming style may influence compressibility}: individual differences disappear by aggregating results from multiple systems from different organisations, written by many developers over the years. The scale and average size of the systems in our benchmark ensure that individual programmer differences cannot play a significant role. Note that \textit{common} programming style differences between programming languages are, and should be, affecting the compression ratios.
	\item \textit{Code structure and maintainability may influence the results}: For example, the amount of duplication in low-maintainable software. To check this, in \cite{Ouwehand2018msc}, the impact of maintainability on compressibility was investigated, but no significant influence was found.
	The amount of modularity and reuse that languages support (either through the language features or the language ecosystems) may influence compressibility, but these are exactly the types of language characteristics that we intend to capture with our language conciseness measurements.
	\item \textit{Library usage and code reuse differ significantly across languages}: This is related to the previous item, and indeed has a big impact on how much code is needed to express a certain amount of \textit{functionality}. However, from a software maintenance perspective, it is not the amount of functionality that is the prime consideration, but the amount of information that a developer has to think about to understand, design, code and debug the software: it is the latter that our approach aims to capture.
	\item \textit{Source code may contain data and natural language}: Indeed, all data constants, both numerical and text, are considered as part of the source code. However, in our benchmark analysis, we have removed lines with only comments and white-space, to ensure the majority of the compressed data is actually source code. We assume that the amount of data constants will not differ significantly across languages, so that its effect is limited when comparing languages.
	\item \textit{No ground truth to validate}: There is no objective reference to determine whether our approach is 'correct' or fully accurate. In our validation section, we have compared to both another measurement-based approach \cite{Berkholz}, and to a large developer survey \cite{MacIver2017}, neither of which can be considered the truth, but both perspectives are relevant. Our motivation is mainly to have an objective approach that uses the same method to determine the conciseness factors for all programming languages.
\end{itemize}
\noindent
Finally, we want to point out the risks of (wrongly) interpreting our language conciseness metrics, as e.g. shown in Table \ref{tbl:cleaning}; these language compression ratio's should not be interpreted as a rating of the 'goodness' of the languages, nor are they directly correlated to how much code is needed to implement functionality (especially libraries and frameworks influence the latter).
So, dear language enthusiasts: your favourite language is not better or worse, just because it has a lower or higher compression ratio value!



\section{Conclusion and Future Work}\label{ref:conclusion}

In this paper, we have investigated an evidence-based approach towards the analysis of \textit{language conciseness}.
The \textbf{first contribution} of this paper is the observation that program conciseness can be defined as the amount of code that is needed to express a single unit of information.
The notion of \textit{Kolmogorov complexity} from information theory provides a theoretical foundation for this approach \textit{and} a practical implementation technique in the form of data compression. This technique has the big advantage that it is \textit{fully independent of the syntax and semantics of the analysed languages}, avoiding the need for language-specific parsing of the source code. The latter is of significant importance for practical application in tooling, since the number of actively used programming languages is high, and actively growing every year.

The \textbf{second contribution} is the detailed description of the approach of data cleaning and in particular our technique for aggregating towards a language-representative value from individual data-points. We present these in sections \ref{ref:approach} and \ref{ref:data_analysis}, including many of the underlying design decisions.

The presentation of the resulting language conciseness metrics, in the form of compression ratios, for a selection of 32 languages is the \textbf{third contribution}; other researchers and practitioners may use that information for further analysis or comparison.
We also share as many details of the actual inputs and results of the data analysis as space permits.

The \textbf{fourth contribution} is our validation of the results, in section \ref{ref:validation}, by (a) Testing that language conciseness (CR) is representative for the systems in our benchmark, and a meaningfully distinctive metric, (b) Comparing to an evidence-based approach\cite{Berkholz} for language expressiveness, and (c) Confirming that our language conciseness values are highly correlated with the results of a large developer survey\cite{MacIver2017}.

As we pointed out in section \ref{ref:relatedwork}, the fundamental ideas underlying this paper are not entirely new. But to the best of our knowledge, there has been no previous publication that documents the actual application of these ideas to a (large) benchmark covering multiple programming languages.


There are several \textbf{next steps} to follow up on the work presented here:
\begin{itemize}
\item We would like to extend our benchmark with an open-source version, this will provide more data-points and open up the data analysis to third parties. For the entries in such a benchmark, we may for example use \cite{Spinellis2020}, where a dataset of enterprise-driven open source system has been collected.
\item There are further promising applications of Kolmogorov Complexity; for example the application of \textit{Normalized Compression Distance}\cite{Cilibrasi2005}, which can provide an indication of the actual amount of code changes over time, that may be more representative than counting the changed lines of code. Normalized Compression Distance has already been applied for plagiarism detection in source code, e.g. in \cite{chen2004,Pribela2019}. 
\item Finally, an interesting exercise would be to apply Normalized Compression Distance to do hierarchical clustering of programming languages (also introduced in \cite{Cilibrasi2005}): that would yield something akin to a programming language genealogy tree\footnote{
	See \texttt{https://www.levenez.com/lang/} for an example that emphasizes the historical evolution.
}.
\end{itemize}

Concluding, we have presented a model for determining the \textit{language conciseness of programming languages} in a systematic and objective, data-based, way. We have presented the approach, detailed analysis steps and quantitative results of applying this model to a large representative benchmark of software applications. This has resulted in language conciseness metrics for over 50 programming languages. We have demonstrated that our metric for language conciseness is strongly correlated with both an alternative analytical approach, and a large scale developer survey.

\vspace{10pt}
\noindent
\textbf{Acknowledgements} We would like to acknowledge Steven Raemaekers, Soerin Bipat and Joost Visser for their contributions to the first application of the ideas in this paper.

%
\bibliographystyle{IEEEtran}
\bibliography{IEEEabrv,library,additional_refs}

\end{document}